\documentclass[prb,superscriptaddress,twocolumn,floatfix]{revtex4-1}
\usepackage[utf8]{inputenc}
\usepackage[T1]{fontenc}
\usepackage{mathptmx}
\usepackage{rotating}
\usepackage{longtable}
\usepackage{listings}
\usepackage{xcolor}
\usepackage{siunitx}
\usepackage{lipsum}
\usepackage{wrapfig}
\usepackage{pgfplots}
\pgfplotsset{compat=1.17}
\usepgfplotslibrary{external}
\pgfplotsset{plot coordinates/math parser=false}
\usetikzlibrary{plotmarks}
\usepgfplotslibrary{colormaps}
\definecolor{codegreen}{rgb}{0,0.6,0}
\definecolor{codegray}{rgb}{0.5,0.5,0.5}
\definecolor{codepurple}{rgb}{0.58,0,0.82}
\definecolor{backcolour}{rgb}{0.95,0.95,0.92}
\lstdefinestyle{mystyle}{
    backgroundcolor=\color{backcolour},   
    commentstyle=\color{codegreen},
    keywordstyle=\color{magenta},
    numberstyle=\tiny\color{codegray},
    stringstyle=\color{codepurple},
    basicstyle=\ttfamily\footnotesize,
    breakatwhitespace=false,         
    breaklines=true,                 
    captionpos=b,                    
    keepspaces=true,                 
    numbersep=5pt,                  
    showspaces=false,                
    showstringspaces=false,
    showtabs=false,                  
    tabsize=2
}

\usepackage{times,amsmath}
\usepackage{epsfig}
\usepackage{color}
\usepackage{longtable}
\usepackage{ulem}
\usepackage{graphicx}
\usepackage{dcolumn}
\usepackage{bm}
\usepackage{bookmark}
\usepackage{tabularx}
\usepackage{hyperref}
\usepackage{multirow}
\usepackage{booktabs}
\usepackage{tocloft}
\usepackage{pgfplots}
\usepackage{pgfplotstable}
\usepackage{tikz}
\hypersetup{colorlinks=true, citecolor=blue, filecolor=blue, linkcolor=blue, urlcolor=blue}

\usepackage[switch, pagewise]{lineno}
\linenumbers\relax 
\nolinenumbers

\begin{document}


\title{Direct Deoxygenation of Phenol over Fe-based Bimetallic Surfaces using On-the-fly Surrogate Models}

\author{Isaac Onyango}
\affiliation{Department of Mechanical Engineering and Engineering Science, University of North Carolina at Charlotte, Charlotte, NC 28223, USA}
\affiliation{North Carolina Battery Complexity, Autonomous Vehicle and Electrification (BATT CAVE) Research Center,
Charlotte, NC 28223, USA}

\author{Qiang Zhu}
\email{qzhu8@charlotte.edu}
\affiliation{Department of Mechanical Engineering and Engineering Science, University of North Carolina at Charlotte, Charlotte, NC 28223, USA}
\affiliation{North Carolina Battery Complexity, Autonomous Vehicle and Electrification (BATT CAVE) Research Center,
Charlotte, NC 28223, USA}

\begin{abstract}
We present an accelerated nudged elastic band (NEB) study of phenol direct deoxygenation (DDO) on Fe-based bimetallic surfaces using a recently developed Gaussian process regression (GPR) calculator. Our test calculations demonstrate that the GPR calculator achieves up to 3x speedup compared to conventional density functional theory (DFT) calculations while maintaining high accuracy, with energy barrier errors below 0.015 eV. Using GPR-NEB, we systematically examine the DDO mechanism on pristine Fe(110) and surfaces modified with Co and Ni in both top and subsurface layers. Our results show that subsurface Co and Ni substitutions preserve favorable thermodynamics and kinetics for both C-O bond cleavage and C-H bond formation, comparable to those on the pristine Fe(110) surface. In contrast, top-layer substitutions generally increase the C-O bond cleavage barrier, render the step endothermic, and result in significantly higher reverse reaction rates, making DDO unfavorable on these surfaces. This work demonstrates both the effectiveness of GRR-accelerated transition state searches for complex surface reactions and provides insights into rational design of bimetallic catalysts for selective deoxygenation.



\end{abstract}
\maketitle
\makeatletter

\section{Introduction}
\label{intro}

Understanding reaction mechanisms of surface-catalyzed processes is crucial for rational catalyst design. The nudged elastic band (NEB) method is widely used to determine minimum energy pathways (MEPs) and identify energy barriers between local minima configurations \cite{NEB-2000, CINEB-2000, Herbo-NEB-2017, JONSSON-1998, caspersen2005finding, sheppard2012generalized, qian2013variable}. In a practical NEB simulation, the method relies on an elastic band of configurations (images) interpolated between local minima. Each image requires force and energy evaluations from first-principles calculations like density functional theory (DFT) during MEP optimization. Since each electronic structure evaluation typically takes tens or even hundreds of CPU minutes, NEB calculations are computationally very demanding. 

In recent years, there have been notable efforts to build machine learning (ML) surrogate models to accelerate NEB calculations \cite{NEB-ML-2016, Jonsson-2017, Jonsson-2019, Prior-Mean-GPR-2022, Prior-Mean-GPR-2024, Schaaf-2023}. These approaches construct models that closely approximate the true potential energy surface (PES), thereby substantially reducing the number of first-principles calculations required during MEP optimization. In these works, machine learning force fields (MLFFs) were typically trained on a great deal of data in order to achieve reliable accuracy. If a new configuration cannot be covered by the training data, one has to retrain the MLFFs in order to achieve a desirable accuracy. In our earlier work, we proposed using a Gaussian process regression (GPR) model to resolve the model update issue for NEB calculations \cite{onyango2025gpr_calculator}. This is essentially a hybrid calculator that integrates both DFT and GPR throughout the NEB calculation, and leverages uncertainties for on-the-fly model learning and efficient updates during the NEB optimization allowing it to dynamically adapt to the evolving potential energy surface. This approach is particularly advantageous as it avoids the need for periodic retraining, which can be computationally expensive and may lead to final MEP inaccuracies if not updated frequently enough. Our approach has demonstrated 3-5 times speedup for Pd$_4$ cluster diffusion on MgO(100) surface and H$_2$S dissociation on transition metal surfaces while maintaining high accuracy \cite{onyango2025gpr_calculator}.

Motivated by the successes on the cases of small molecules adsorbed on surfaces, we aim to extend this approach to study systems consisting of larger molecules that are generally considered too computationally demanding for routine DFT-based NEB simulations. In this work, we explore the deoxygenation (DDO) of phenol on Fe-based bimetallic surfaces, which presents additional challenges due to the structural rearrangements and rotational flexibility associated with bond breaking and formation in such a larger molecules on the surface. From a technological perspective, DDO is a particularly attractive hydrodeoxygenation (HDO) pathway, as it yields aromatic products with minimal hydrogen consumption. This offers the potential for HDO under ambient conditions, presenting a promising alternative to traditional HDO processes. However, the DDO of phenolics is challenging due to the high energy barrier associated with the C-O bond cleavage \cite{nie2014selective}. 

Transition metal bimetallic catalysts enhance catalytic activity and selectivity for aromatic production (e.g., benzene, toulene and xylene) compared to monometallic catalysts \cite{nie2014selective, Lui-2018, kordouli2017hdo, huynh2014hydrodeoxygenation, yang2017geometric, zhao2012comparison, do2012bimetallic,liu2018hydrodeoxygenation}. Fe-based catalysts demonstrate up to 90\% selectivity but suffer from oxidative deactivation. Alloying Fe with other transition metals improves catalyst stability. For example, Hong et al. \cite{C-Pd-Fe-2013, Pd-Fe-2014,hong2017stabilization} showed that alloying with noble metals like Pt, Pd, Rh and Ru reduces oxidative deactivation while maintaining high aromatic selectivity, especially in vapor phase conditions.

Most DFT studies of HDO on bimetallic catalysts focus on single-atom alloy (SAA) models that emulate surfaces with low local concentrations of the secondary metal \cite{nie2021effect, zhou2019hydrodeoxygenation, li2021computational, jia2019effect}. However, high local coverages of the second metal can significantly influence surface energetics and reaction mechanisms. For example, studies of oxygen reduction reaction (ORR) on Pt$_3$Ni(111) surfaces found that multilayer Pt skin terminations exhibit higher activity \cite{cao2015rational,wang2011design}. Similarly, Jiang et al. \cite{fufural-M-Pt111} showed that Co, Fe and Ni terminated surfaces enhance furfural HDO activity compared to pure Pt surfaces, though their DFT calculations were limited to reaction energies without full minimum energy pathways.

In this work, we focus on Fe-based bimetallic catalysts, specifically Fe(110) surfaces with Co and Ni incorporated via top-layer and subsurface-layer substitution. Co and Ni were selected as they are show potential for enhanced HDO activity and are more earth-abundant compared to noble metals, making them a more sustainable and cost-effective alternative. We employ a GPR calculator to accelerate NEB calculations for exploring the DDO mechanism of phenol on these surfaces. After validating the GPR model on pristine Fe(110), where it achieves 3x speedup while maintaining high accuracy, we systematically investigate the bimetallic systems. Our results reveal that subsurface alloying with Co or Ni maintains favorable DDO energetics similar to pristine Fe(110), while top-layer alloying generally leads to less favorable reaction pathways.

\section{Computational Methodology}

\subsection{Standard DFT-NEB Setup}
DFT calculations were performed using the Vienna \textit{ab initio} simulation package (VASP) \cite{VASP01-1996, VASP02-1996, VASP03-1993}. Core electrons and electron-electron exchange correlation effects were treated using the projector augmented wave (PAW) \cite{PAW-potentials-1999} method and optB88-vdW functional \cite{optb88}, respectively. Spin polarization and dipole correction (applied perpendicular to the surface) were included. Valence electrons were modeled using a plane-wave basis set with a 400 eV cutoff energy. The Methfessel-Paxton smearing method \cite{Methfesel-1989} with 0.1 eV width was used. Calculations were considered converged when energy differences and forces were below $10^{-4}$ eV and 0.03 eV/\AA~, respectively. Surfaces were modeled using 4-layer slabs with 15 \AA~ vacuum spacing to prevent interaction between periodic images. The top two layers were allowed to relax while the bottom two were fixed. The Brillouin zone was sampled using a $\Gamma$-centered $2 \times 2 \times 1$ Monkhorst-Pack k-point grid.

Transition states (TS) were identified using the climbing image nudged elastic band (CI-NEB) method \cite{CINEB-2000}. NEB calculations were considered converged when forces on all images were below 0.075 eV/\AA~, using the FIRE algorithm \cite{FIRE-2006} implemented in the atomic simulation environment (ASE) \cite{ASE-2017}. TS structures were verified by calculating vibrational frequencies using central finite differences with a 0.015 \AA~ step size. Within the harmonic approximation, each TS was confirmed by the presence of exactly one imaginary frequency \cite{trygubenko2004doubly}. For vibrational calculations, the surface was fixed while adsorbates were allowed to relax.

\subsection{The GPR-NEB Setup}
A Gaussian process is a probability distribution over functions that fit a collection of points \cite{bishop2006pattern}. It is characterized by a mean function, typically assumed to be zero, and a covariance function (kernel) that defines the correlation between points. Here the kernel is the radial basis function defined as:
\begin{equation}
    k(x_i, x_j) = \sigma_m^2 \exp\left(-\frac{(x_i - x_j)^2}{2l^2}\right)
\end{equation}
where $\sigma_m$ and $l$ are the hyperparameters that control the magnitude of the covariance function and length scale, respectively. For a set of sample data \{$\boldsymbol{x}, Y$\}, where $\boldsymbol{x}$ represents the input vector and $Y$ is the vector of corresponding observations, the covariance matrix $\boldsymbol{C}$ is constructed as:
\begin{equation}
     \boldsymbol{C}_{mn}=\boldsymbol{C}(x_m, x_n) = k(x_m, x_n) + \beta \boldsymbol{\delta}_{mn},
\end{equation}
where $\beta$ is the noise variance, and $\boldsymbol{\delta}_{mn}$ is the Kronecker delta function. For $N$ samples, $\boldsymbol{C}$ is a square matrix of $N \times N$. Each sample value can be considered as the linear combination of these covariances.
\begin{equation}
    \boldsymbol{Y_m} = \sum_{i=1}^N \boldsymbol{G}_{\alpha i} \boldsymbol{C}(\boldsymbol{x}_m, \boldsymbol{x}_i)
\end{equation}
Hence, one only needs to determine $\boldsymbol{G}_{\alpha}$ from the previous training data. In matrix form, $\boldsymbol{Y}=\boldsymbol{C}\boldsymbol{G}_\alpha$. For a new point $\boldsymbol{x}_{N+1}$, the covariance vector $\boldsymbol{C}_{N+1}$ is extended as:
\begin{equation}
    \boldsymbol{C}_{N+1} = 
    \begin{pmatrix}
    \boldsymbol{C}_N & \boldsymbol{k} \\
    \boldsymbol{k}^T & c      
    \end{pmatrix}
\end{equation}
where $c = k(x_{N+1}, x_{N+1}) + \beta$. And the vector $\boldsymbol{k}$ has elements $k(x_n, x_{N+1})$ for $n = 1, \cdots, N$. The prediction output  and variance for the new point are given by:
\begin{equation}\label{pred}
    \boldsymbol{Y_{N+1}} = \boldsymbol{C}_{N+1} \boldsymbol{G_\alpha}
\end{equation}
\begin{equation}\label{std}
    \boldsymbol{\sigma}^2_{N+1} = c - \boldsymbol{k}^T \boldsymbol{C}_N^{-1} \boldsymbol{k}
\end{equation}

The predictive performance of the GPR is improved by optimizing the hyperparameters $\sigma_m$, $l$, and $\beta$ using the maximum likelihood estimation method. For our application, each input structure is represented by a vector of structural descriptors and their partial derivatives. The descriptors encode the local atomic environment and must respect translation, rotation and permutation symmetry. In our implementation, we used the SO(3) descriptors derived from the power-spectrum of spherical harmonic expansion coefficients \cite{yanxon2020pyxtalff, zagaceta2020}. Below we summarize the workflow of the NEB calculations with the GPR calculator. For more mathematical and implementation details, see our previous work \cite{onyango2025gpr_calculator}. 
\begin{enumerate}
    \item Given input initial and final states, an initial trajectory was generated by interpolation between them. 
    And these images together with the initial and final states, along with their DFT-calculated energies and forces served as initial training data for the GPR model.
    \item Within each NEB optimization step, the GPR calculator predicts the energy and forces along with the corresponding uncertainties for each image. If the uncertainty in the predictions exceeds a predefined threshold, the model calls the VASP calculator to compute the DFT energy and forces for the NEB calculation and simultaneously adds the simulation results to the training database and then updates the model. This process continues until the NEB calculation converges. Here, we set a threshold for the maximum uncertainty in energy and forces to $\sigma_E$ = 0.05 eV/structure for energy and $\sigma_F$ = 0.075 eV/\AA~ for forces, respectively. 
\end{enumerate}


\subsection{Rate Constants Estimation}
In addition to MEPs, we estimated of the forward rate constants and equilibrium constants using the transition state theory (TST) \cite{QHTST,laidler1987chemical}. The forward rate constant $k_f$ is calculated using the following equation:
\begin{equation}
k_{\text{f}} = \frac{k_B T}{h} \cdot \frac{Q^\ddagger_{vib}}{Q_{vib}} \cdot e^{-\frac{E^\ddagger_{\text{a}}}{k_B T}}
\end{equation}
where $k_B$ is the Boltzmann constant, $T$ is the temperature, $h$ is Planck's constant, $Q^\ddagger_{vib}$ and $Q_{vib}$ are the vibrational partition functions of the transition state and reactant state, respectively, and $E^\ddagger_{\text{a}}$ is the activation energy barrier. The vibrational partition function is calculated using equation:
\begin{equation}
Q_{\text{vib}} = \prod_{i=1}^{N} \frac{e^{-h\nu_i / 2k_B T}}{1 - e^{-h\nu_i / k_B T}}
\end{equation}
where $\nu_i$ is the vibrational frequency of the $i$-th mode and $N$ is the number of vibrational modes. The reversed rate constant $k_r$ is calculated using similar equation. The equilibrium constant $K_\text{eq}$ is calculated using the equation:
\begin{equation}
K_\text{eq} = \frac{k_f}{k_r} 
\end{equation}

In this work, both DFT-NEB and GPR-NEB calculations were run in a single compute node of AMD EPYC 9654P 96-Core Processor with 2.40 GHz in our in-house computing cluster.

\section{Results and Discussion}
\label{results}
In this section, we first validate the GPR calculator's performance by comparing its results against standard DFT calculations for phenol DDO on pure Fe(110). After establishing the accuracy and computational efficiency of our approach, we systematically investigate how surface modification impacts the reaction mechanism by examining both top-layer and sub-layer substitution of Fe with Co and Ni. The reaction pathways and energetics are analyzed in detail for each surface configuration. Finally, we compare kinetic parameters across all surfaces to evaluate their catalytic performance and identify promising catalyst designs.

\subsection{GPR Calculator Performance on a Pure Fe(110) surface}

\begin{figure}[h] 
    \centering
    \includegraphics[width=0.48\textwidth]{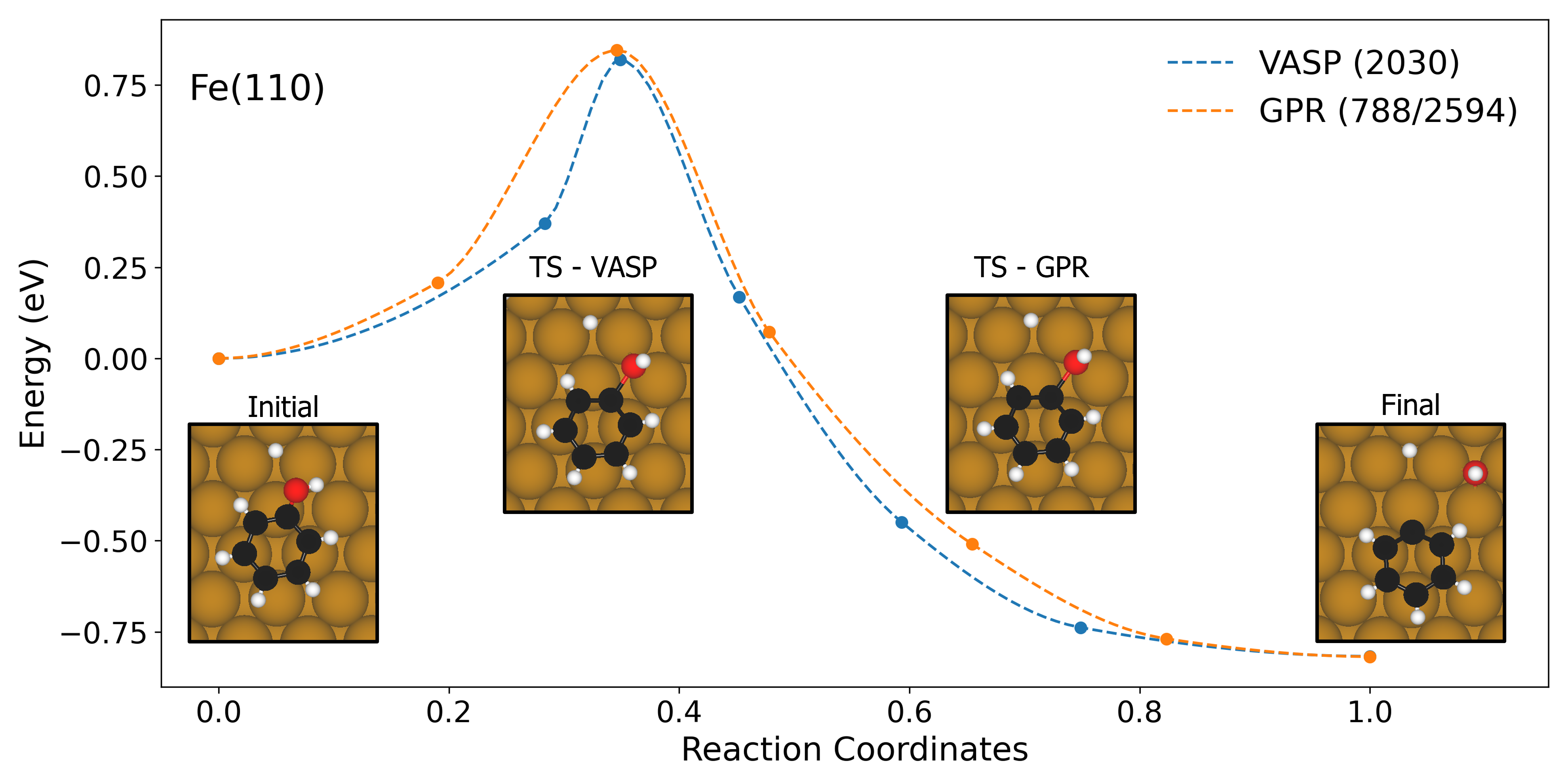}
    \caption{The simulated MEP of deoxygenation of phenol on the Fe(110) surface from both the GPR and pure VASP calculators. The representative structures along the transition path are also shown in the inset. Brown, white and red spheres represent Fe, O and C atoms, respectively.}
    \label{fig1}
\end{figure}

We began with test calculations to assess the accuracy and computational efficiency of the GPR calculator by studying the C-O bond scission of phenol on Fe(110) surface. As shown in Figure \ref{fig1}, the MEPs calculated using both the GPR and VASP are nearly identical. The difference in the energy barrier is negligible: 0.813 eV from the GPR compared to 0.827 eV from the pure VASP calculations. Additionally, the TS structures are nearly indistinguishable. In both cases, the OH resides near a bridge site, and the C-O bond is significantly elongated from 1.39 \AA~ to 1.88 \AA~ for the TS from GPR and 1.96 \AA~ for the TS from VASP. The error in the energy barrier is only 0.014 eV, which is well within the acceptable range for NEB calculations. 

\begin{figure*}[htbp]
    \centering
    \includegraphics[width=0.85\textwidth]{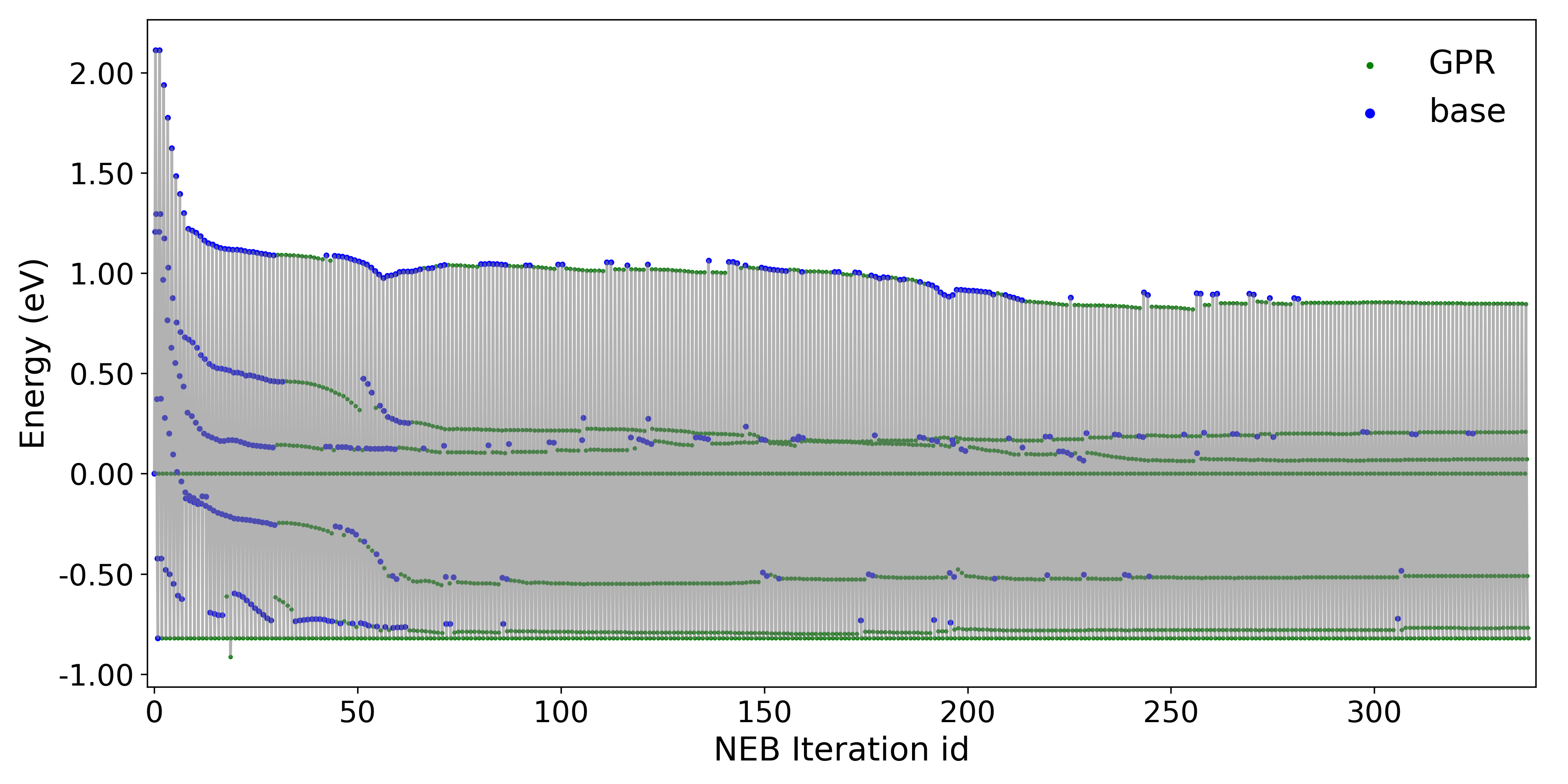}
    \caption{The usage of GPR and base calls during the MEP optimization associated with the GPR-NEB simulation mentioned in Fig. \ref{fig1}. In this simulation, each NEB iteration includes 7 consecutive images.}
    \label{fig2}
\end{figure*}


In terms of computational efficiency, the GPR calculator completed the MEP optimization in 72 hours, requiring 788 DFT calls and 2594 GPR calls. Using the same setup, it costs a total of 216 hours based on the VASP calculator with 2030 DFT calls. This demonstrates a 3 times speedup of GPR over the pure VASP-NEB approach. 

To understand the acceleration mechanism, we plot Figure \ref{fig2} to analyze the distribution of GPR and VASP calls during MEP optimization. The first 40 iterations rely predominantly on VASP calculations to build an accurate initial GPR model. Between iterations 40-60, most force and energy evaluations use the GPR calculator, except for the 6th image which requires VASP calculations due to significant structural changes. From iterations 50-210, optimization is dominated by GPR calls, with VASP calculations concentrated near the transition state where major structural rearrangements occur. After iteration 210, the optimization proceeds almost entirely through GPR, requiring only occasional VASP calls for structural corrections. This nonuniform pattern of VASP calls demonstrates how the GPR calculator's on-the-fly model updates, guided by uncertainty estimates, enable dynamic adaptation to the evolving potential energy surface while maintaining accuracy. As long as the GPR is well trained, it takes only 10-50 seconds to predict the atomic forces and energies to drive the NEB optimization, while each DFT calculation typically takes 45-55 minutes for a surface model consisting of 78 atoms in the unit cell.

\begin{figure}[htbp] 
    \centering
    \includegraphics[width=0.48\textwidth]{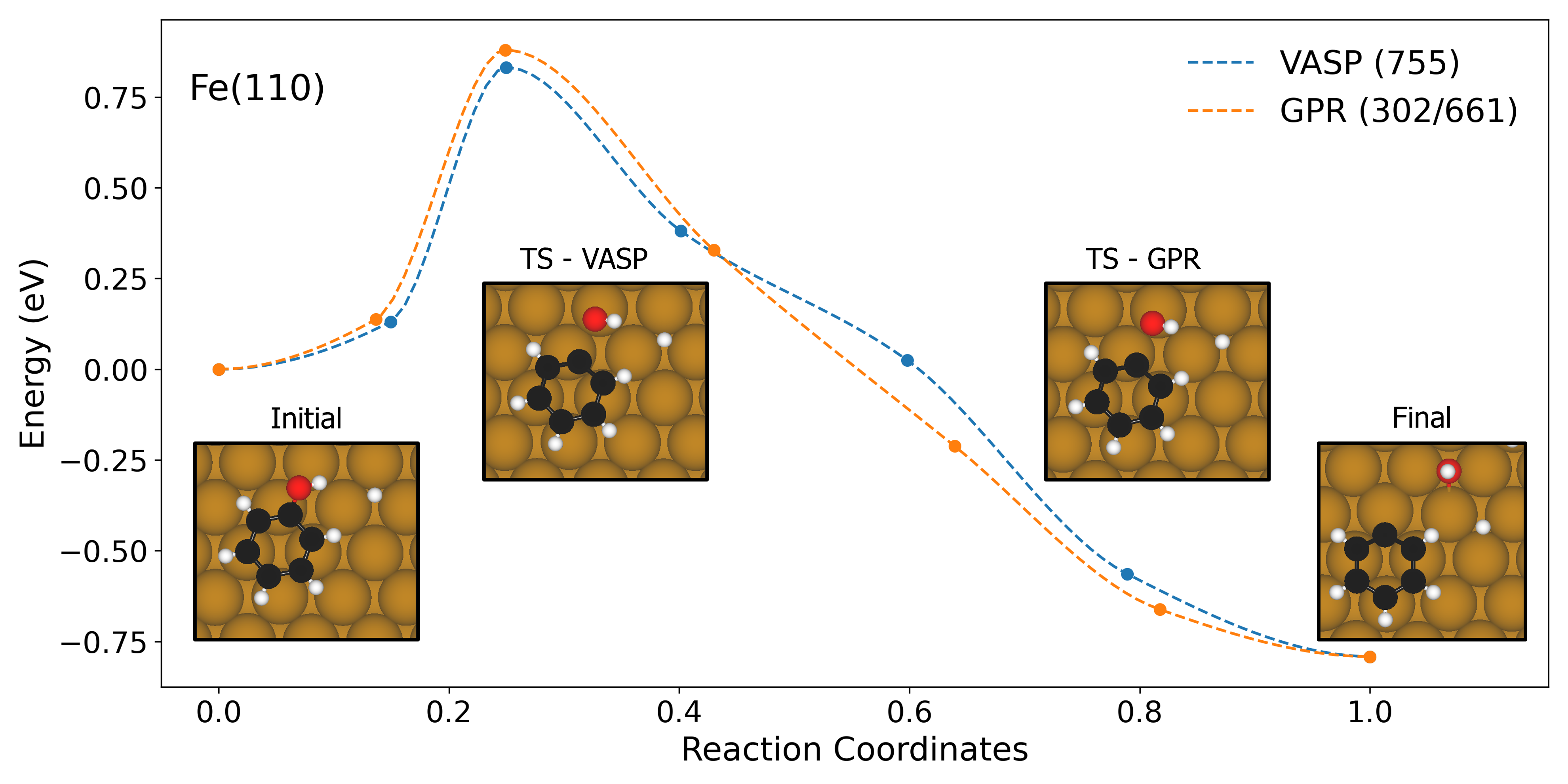}
    \caption{The simulated MEP of deoxygenation of phenol on the Fe(110) surface from both the GPR and pure VASP calculators. The representative structures along the transition path are also shown in the inset. Brown, white and red spheres represent Fe, O and C atoms, respectively.}
    \label{fig3}
\end{figure}

It is important to note that the GPR acceleration efficiency depends on the NEB path quality. In another simulation with a slightly different initial path (see Figure \ref{fig3}), the GPR calculator achieved a more modest 1.8x speedup - completing the MEP optimization in 2.5 days (302 VASP calls, 661 GPR calls) compared to 4.5 days for pure VASP (755 DFT calls). The key difference between these paths lies in the molecular motions during the transition. In the previous case (Figure \ref{fig1}), the TS formation involved both OH group rotation during C-O bond scission and aromatic ring translation/rotation. In contrast, such complex rotational motions were minimal in this example. When present, these additional rotational degrees of freedom can increase the number of iterations needed for NEB convergence. This suggests that optimal performance requires both GPR acceleration and careful selection of the initial reaction path to minimize unnecessary molecular rotations.

\subsection{Phenol DDO on Fe-based Bimetallic Surfaces}

Having confirmed the validity of the GPR calculator for studying phenol DDO on Fe(110), we proceeded to investigate the impact of surface doping. For a systematic understanding, we considered three types of surface models: (i) pure Fe(110); (ii) bimetallic surfaces with the top Fe(110) layer substituted by Ni or Co; and (iii) bimetallic surfaces with the sub-layer substituted by Ni or Co. For each surface, we studied both the C-O bond cleavage and C-H bond formation steps, resulting in a total of 10 independent NEB simulations. Given that each pure VASP NEB simulation would take 4-9 days, using the GPR calculator allowed us to complete each simulation in just 2-3 days while maintaining accuracy.

The reaction mechanism involves two key steps, starting with phenol and hydrogen coadsorbed horizontally on the surface. First, the C-O bond cleaves to form adsorbed phenyl (C$_6$H$_5$) and hydroxyl species, along with the remaining hydrogen. Subsequently, C-H bond formation occurs as the hydrogen binds to the phenyl group to produce benzene. Below we analyze these reaction steps in detail for each surface configuration.

\subsubsection{Phenol DDO on pure Fe(110)}

\begin{figure}[htbp]
    \centering
    \includegraphics[width=0.48\textwidth]{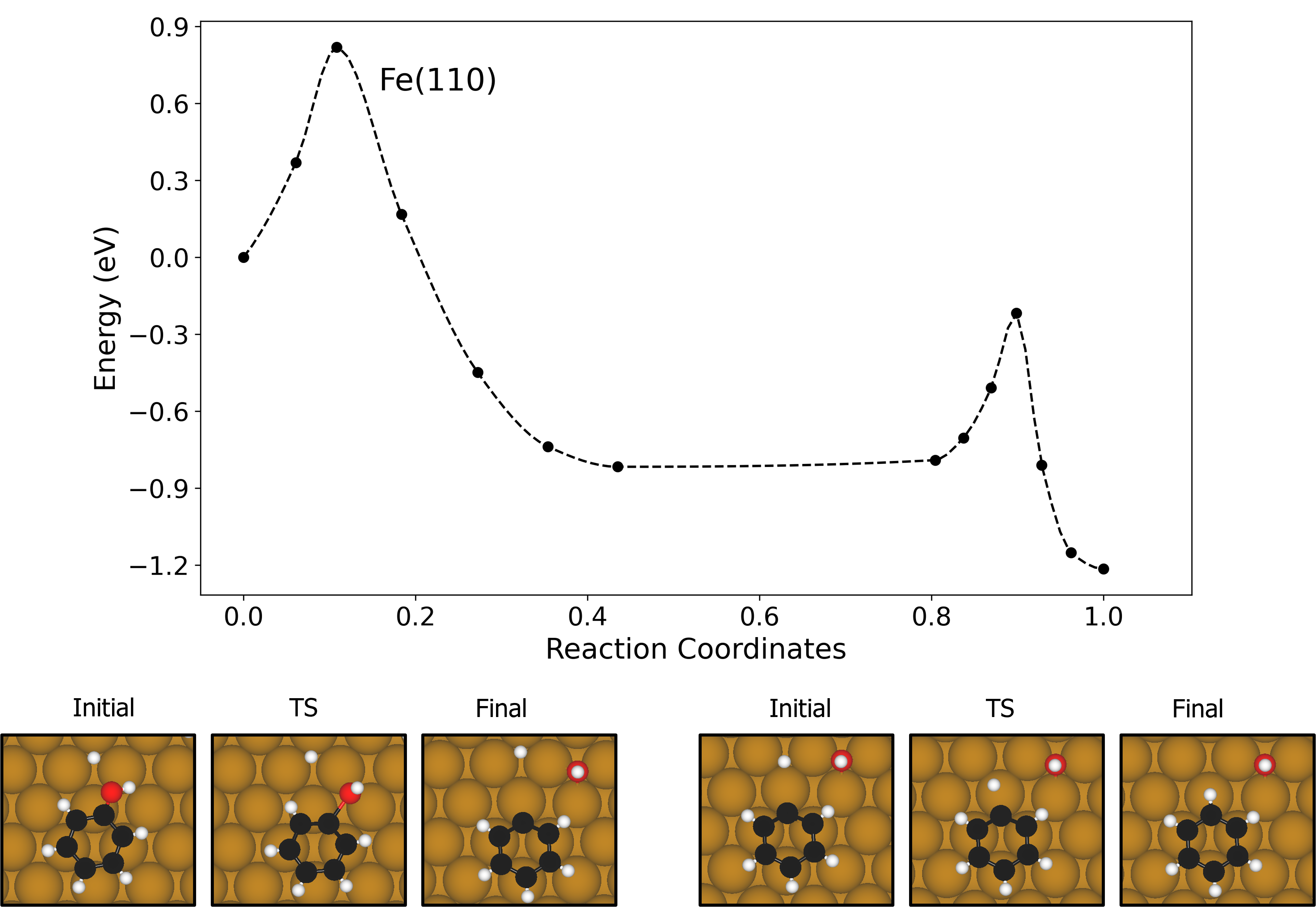}
    \caption{The simulated MEPs of the deoxygenation and hydrogenation steps on the pure Fe(110) surfaces. The initial, transition state (TS) and final structures the elementary steps are shown below the MEPs. Brown, white and red spheres represent Fe, O and C atoms, respectively.}
    \label{fig-pure-fe}
\end{figure}

The reaction on the pure Fe surface is summarized in Figure \ref{fig-pure-fe}. The C-O bond cleavage proceeds via a slight translation of the phenol molecule, accompanied by rotation of the OH group to a vertical orientation, with the oxygen atom interacting directly with a bridge site on the surface. At the transition state (TS), the C-OH bond is significantly elongated, increasing from 1.39 \AA~ to 1.88 \AA~. During C-H bond formation, the hydrogen atom migrates from a three-fold hollow site toward a top site at the TS, rising by approximately 0.50 \AA~ above its initial position near the surface at the TS. This overall reaction is highly exothermic, with a reaction energy of -1.22 eV. Both the C-O bond cleavage and C-H bond formation steps are exothermic (-0.82 eV and -0.41 eV, respectively) and exhibit moderate activation barriers of 0.82 eV and 0.57 eV. These results suggest that DDO is readily proceeds on the pure Fe surface, with the C-O cleavage is the rate limiting step. 

\subsubsection{Phenol DDO on the top-layer substitution}

\begin{figure}[htbp]
    \centering
    \includegraphics[width=0.48\textwidth]{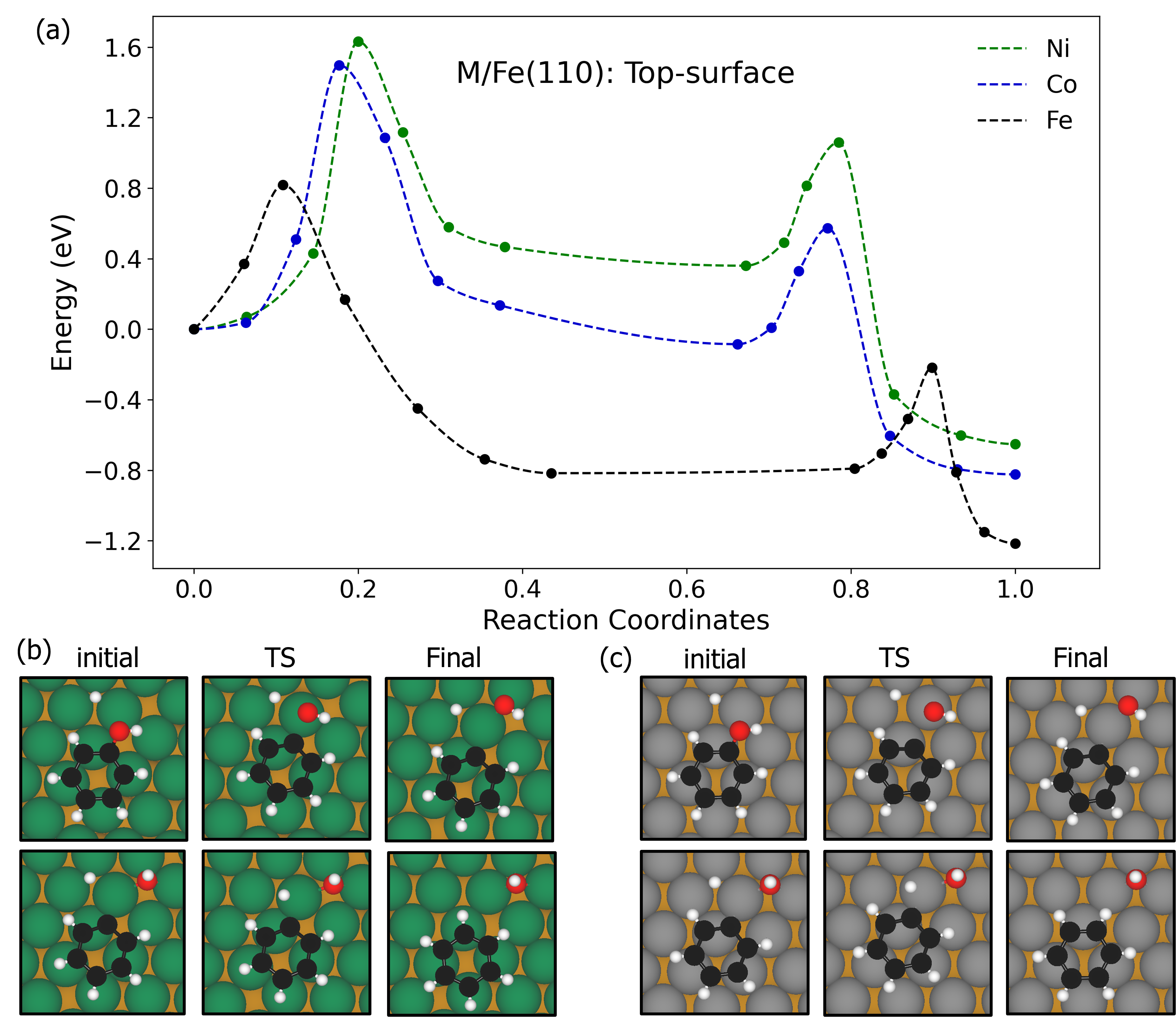}
    \caption{The simulated MEPS of the deoxygenation and hydrogenation steps on Co and Ni top-layer substitution surfaces. (b) The initial, transition state (TS) and final structures for the C-O bond cleavage (top row) and C-H formation (bottom row) steps on the Ni top-surface. (c) The initial, transition state (TS) and final structures for the same steps for the Co top-surface. Brown, green, gray, white and red spheres represent Fe, Ni, Co, O and C atoms, respectively.}
    \label{fig5}
\end{figure}

The MEPs on top-layer substitution surfaces are shown in Figure \ref{fig5}. The C-O cleave proceeds via slight translation/rotation of the phenol accompanied by migration of the OH group, with minimal rotation, toward the top site at the TS. The C-O bond is significantly elongated from about 1.38 \AA~ to 2.05 \AA~. The C-H bond formation occurs via a similar mechanism, with the hydrogen migrating from a three-fold hollow site to a top site, rising by approximately 0.77 \AA~ above its initial position at the TS. 

The overall reaction is exothermic, with the reaction energy of -0.70 eV for Co and -0.66 eV for Ni surfaces. However, the C-O bond cleavage on these surfaces is energetically unfavorable, exhibiting high activation barriers of 1.14 eV and 1.47 eV, and endothermic reaction energies of 0.09 eV and 0.40 eV, for Co and Ni surfaces, respectively. In contrast, C-H bond formation is more favorable, lower activation barriers of 0.62 eV and 0.58 eV, and  exothermic reaction energies of -0.45 eV and -0.40 eV, for Co and Ni surfaces, respectively. These results indicate that C-O bond cleavage is the rate limiting step on these surfaces. Although the overall reaction is exothermic, the unfavorable energetics of the initial cleavage suggest that DDO is not readily facilitated on these surfaces.

\subsubsection{Phenol DDO on the sub-layer substitution}

\begin{figure}[htbp]
    \centering\includegraphics[width=0.48\textwidth]{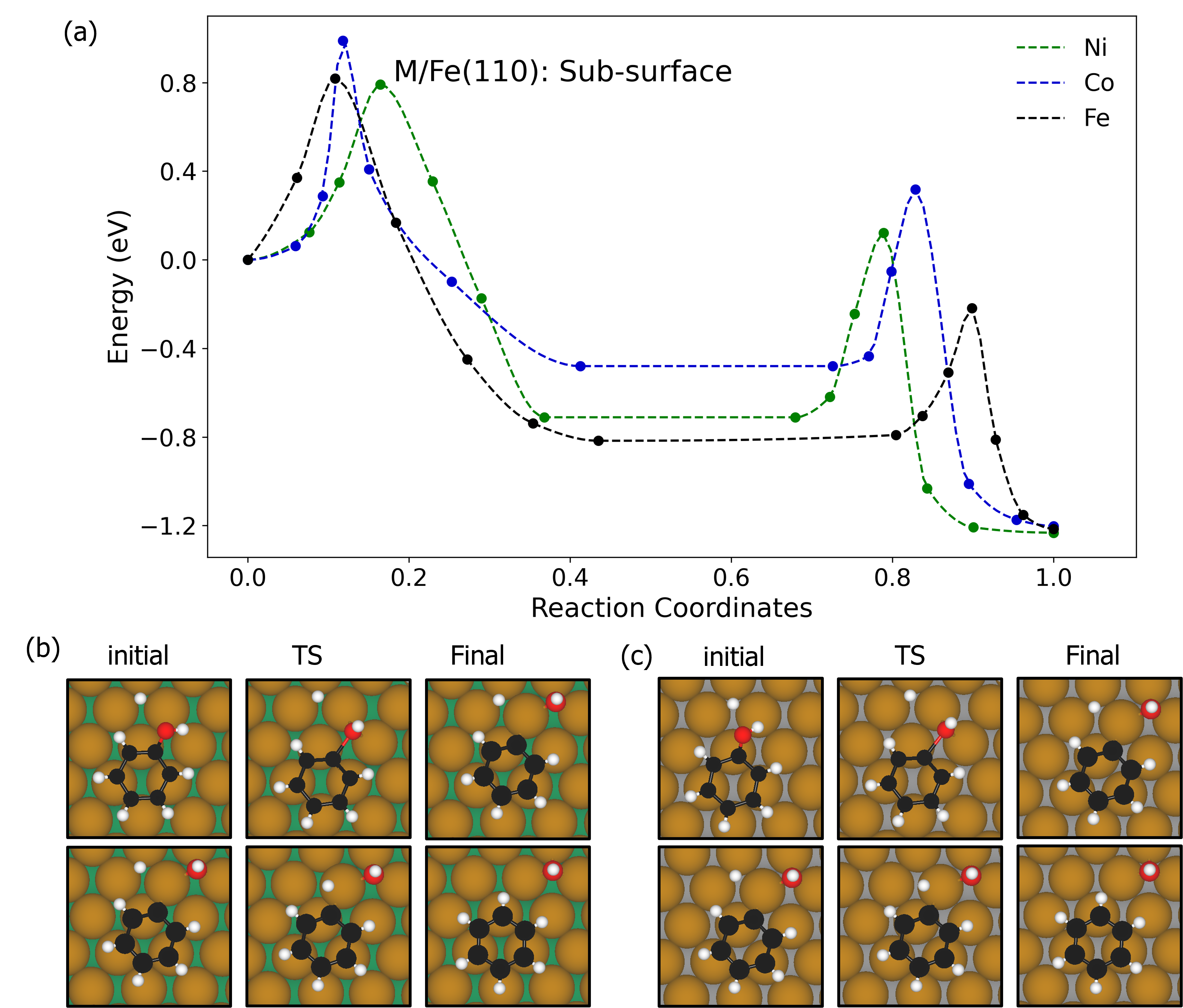}
    \caption{The simulated of the deoxygenation and hydrogenation steps on the Co and Ni sub-layer substitution surfaces. (b) The initial, transition state (TS) and final structures for the C-O bond cleavage (top row) and C-H formation (bottom row) steps on the Ni sub-surface. (c) The initial, transition state (TS) and final structures for the same steps for the Co sub-surface. Brown, green, gray, white and red spheres represent Fe, Ni, Co, O and C atoms, respectively.}
    \label{fig6}
\end{figure}

Figure \ref{fig6} summarizes the overall reaction on the sub-layer substitution surfaces. Similar to the pure Fe surface, the C-O cleavage also proceeds via a slight translation of the phenol molecule, accompanied by rotation of the OH group to a vertical orientation, with the oxygen atom interacting directly with a bridge site on the surface at the TS. TS structures are similar to that of the pure Fe surface, with the C-O bond elongated from 1.39 \AA~ to 1.92 \AA~ for Co and 1.38 to 1.88 \AA~ for Ni surfaces. The C-H bond formation also occurs similarly to the pure Fe surface, with the hydrogen atom migrating from a three-fold hollow site toward a top site at the TS, rising by approximately 0.52 - 0.79 \AA~ above its initial position near the surface at the TS.

The overall reactions are highly exothermic with total energies of -1.20 eV for Co and -1.23 eV for Ni surfaces. Both the C-O bond cleavage and C-H bond formation steps exhibit favorable energetics. The C-O bond cleavage has moderate activation barriers of 0.99 eV and 0.79 eV with exothermic reaction energies of -0.48 eV and -0.71 eV for Co and Ni surfaces, respectively. Similarly, the C-H bond formation shows moderate activation barriers of 0.80 eV and 0.83 eV with exothermic reaction energies of -0.72 eV and -0.52 eV for Co and Ni surfaces, respectively. The comparable activation barriers between the two steps suggest they compete for being rate-limiting. Overall, the moderate barriers and exothermic nature indicate that DDO is readily facilitated on these sub-layer substituted surfaces.

\subsubsection{Comparisons of different surface models}

To further assess the kinetics of the DDO of phenol on these surfaces, we estimated the forward rate constants and equilibrium constants at \(350^\circ\mathrm{C}\) and \(450^\circ\mathrm{C}\) using the calculated activation and reaction energies, along with the vibrational frequencies of the initial, transition, and final state structures (see Table \ref{tab1}). These temperatures have been shown experimentally to be efficient for HDO of phenolics \cite{C-Pd-Fe-2013}. 

On the top-layer substitution surfaces (Co\_top and Ni\_top), the hydrogenation step is thermodynamically favorable. However the equilibrium constants for the initial C-O cleavage step indicate strong preference for the reverse reaction. On the Ni\_top surface the reverse reaction dominates, with K$_{eq}$ = \(8.1 \times 10^{-4}\) at \(350^\circ\mathrm{C}\) and \(2.0 \times 10^{-3}\) at \(450^\circ\mathrm{C}\), which indicate 3 - 4 orders of magnitude faster than the forward reaction. On the Co\_top surface, the reverse reaction also dominates (K$_{eq}$ = \(7.6 \times 10^{-2}\) at \(350^\circ\mathrm{C}\) and \(8.5 \times 10^{-2}\) at \(450^\circ\mathrm{C}\)), approximately 2 orders faster than the forward reaction. This suggests that the C-O cleavage is very unlikely on these surfaces. The unfavorable C-O cleavage kinetics implies that top layer alloying with Ni and Co may not be beneficial for DDO reactions.

On the other hand, the sub-layer substitution surfaces (Co\_sub and Ni\_sub) appears to be feasible strategy. It offers C-O cleavage kinetics comparable to pure Fe and potentially improving benzene yield through enhanced C-H bond formation. On the Co\_sub surface. While C-O cleavage is not enhanced compare to pure Fe, it still proceeds readily. On Co\_sub, the forward reaction is about 3-4 orders of magnitude faster (K$_{eq}$ = \(1.4 \times 10^{3}\) at \(350^\circ\mathrm{C}\) and \(4.7 \times 10^{2}\) at \(450^\circ\mathrm{C}\)) than the reverse reaction and approximately 4-5 orders of magnitude fater on Ni\_sub (K$_{eq}$ = \(1.3 \times 10^{5}\) and \(2.4 \times 10^{4}\) at the same temperatures). The C-H formation, however, is enhanced on these surfaces. For Co\_sub, the forward reaction is 4-5 orders of magnitude faster (K$_{eq}$ = \(2.2 \times 10^{5}\) at \(350^\circ\mathrm{C}\) and \(4.7 \times 10^{4}\) at \(450^\circ\mathrm{C}\)) than the reverse reaction compared to pure Fe (K$_{eq}$ = \(6.6 \times 10^{2}\) and \(3.0 \times 10^{2}\)). Ni\_sub shows moderate enhancement, with the forward reaction rates of 3-4 orders magnitude faster (K$_{eq}$ = \(5.0 \times 10^{3}\) and \(1.8 \times 10^{3}\)). 

\begin{table}[htbp]
\begin{center}
\caption{The calculated forward rate constants (k$_f$) and equilibrium constants (k$_{eq}$) for the C-O cleavage and C-H formation reactions at 350 and 450 \(^{\circ}\mathrm{C}\).}
\label{tab1}
\begin{tabular}{cc cc cc}

\hline\hline
\noalign{\vskip 0.1cm}
Surface & Reaction & \multicolumn{2}{c}{T = 350 \(^{\circ}\mathrm{C}\)} & \multicolumn{2}{c}{T = 450 \(^{\circ}\mathrm{C}\)} \\ \cline{3-6}
        &       & k$_f$ (s$^{-1}$) & K$_{eq}$ & k$_f$ (s$^{-1}$) & K$_{eq}$ \\
\noalign{\vskip 0.1cm}
\midrule
\noalign{\vskip 0.25cm}
\multirow{2}{*}{Fe} & \multirow{1}{*}{C-O cleavage} & \num{2.9e6} & \num{9.0e06} & \num{2.5e07} & \num{1.1e+06} \\
                   & \multirow{1}{*}{C-H formation}  & \num{7.3e08} & \num{6.6e02} & \num{3.2e09} & \num{3.0e02} \\
\noalign{\vskip 0.25cm}
\multirow{2}{*}{Co\_top} & \multirow{1}{*}{C-O cleavage} & \num{7.7e03} & \num{7.6e-02} & \num{1.3e05} & \num{8.5e-02} \\
                   & \multirow{1}{*}{C-H formation}  & \num{8.4e07} & \num{2.2e+05} & \num{4.7e08} & \num{4.7e+04} \\
\noalign{\vskip 0.25cm}
\multirow{2}{*}{Co\_sub} & \multirow{1}{*}{C-O cleavage} & \num{1.7e+06} & \num{1.4e+03} & \num{1.6e07} & \num{4.7e02} \\
                   & \multirow{1}{*}{C-H formation}  & \num{2.9e+06} & \num{2.2e+05} & \num{2.0e07} & \num{4.7e04} \\
                   \noalign{\vskip 0.25cm}
\multirow{2}{*}{Ni\_top} & \multirow{1}{*}{C-O cleavage} & \num{4.1e+01} & \num{8.1e-04} & \num{1.6e03} & \num{2.0e-03} \\
                   & \multirow{1}{*}{C-H formation}  & \num{1.4e07} & \num{2.0e08} & \num{1.1e08} & \num{1.6e07} \\
\noalign{\vskip 0.25cm}
\multirow{2}{*}{Ni\_sub} & \multirow{1}{*}{C-O cleavage} & \num{9.4e+07} & \num{1.3e05} & \num{5.2e08} & \num{2.4e04} \\
                   & \multirow{1}{*}{C-H formation}  & \num{5.0e06} & \num{5.0e03} & \num{4.3e07} & \num{1.8e03} \\

\noalign{\vskip 0.25cm}
\hline\hline                
\end{tabular}
\end{center}
\vspace{-0.5em}
\end{table}

Hence, Co\_sub and Ni\_sub surfaces exhibit similar performance to the pure Fe surface. The overall reaction remains highly exothermic, and both elementary steps are thermodynamically and kinetically favorable. This is especially true for the Ni\_sub surface, where both the activation barriers and reaction energies are within approximately 0.2 eV of those on the pure Fe surface. While equilibrium constants indicate that C-O cleavage proceeds more readily on the pure Fe surface, the difference is minimal for Ni\_sub and more pronounced for Co\_sub. These results suggest that subsurface alloying, particularly with Ni, maintains efficient C-O cleavage and substantially improves C-H bond formation, potentially enhancing benzene yield in DDO processes.




\section{Conclusion}
\label{conclusion}

We have demonstrated the effectiveness of the GPR calculator in accelerating NEB calculations for large molecules on surfaces, using phenol on Fe-based bimetallic surfaces as a case study. The GPR calculator significantly reduces the computational cost while maintaining accuracy, achieving up to 3 times speedup compared to pure DFT calculations. Given the high selectivity of Fe-based catalysts for aromatic production and their susceptibility to oxidative deactivation, our results suggest that if Co or Ni are introduced to enhance stability through alloying, subsurface incorporation represents a promising strategy that preserves the catalytic selectivity of Fe-based catalysts. This modification preserves favorable kinetics for C-O bond cleavage and C-H bond formation, comparable to those on pure Fe surfaces. In contrast, top-layer alloying leads to unfavorable C-O cleavage kinetics, indicating it may hinder DDO performance. Overall, these findings underscore the potential of GPR-based methods for efficiently studying complex surface reactions involving larger molecular systems.

\section*{Acknowledgments}
This research was sponsored by the U.S. Department of Energy, Office of Science, Office of Basic Energy Sciences and the Established Program to Stimulate Competitive Research (EPSCoR) under the DOE Early Career Award No. DE-SC0024866. The computing resources are provided by ACCESS (TG-DMR180040). 

\bibliographystyle{elsarticle-num}
\bibliography{2025-cpc-gpr}
\end{document}